\documentclass[a4paper,11pt]{article}
\usepackage{slashed}

\usepackage{braket}

\pdfoutput=1 

\usepackage{jheppub} 

\usepackage[T1]{fontenc} 

\usepackage{xcolor} 

\title{Topological Structure of Infrared QCD}

\author[a,1]{J . Gamboa,\note{Corresponding author}}
\affiliation[a]{Departamento de Física, Universidad de Santiago de Chile,
\\Av. Víctor Jara 3493, Santiago 9170020, Chile}

\emailAdd{jorge.gamboa@usach.cl}

\abstract{
We investigate the infrared structure of QCD within the adiabatic approximation,
where soft gluon configurations evolve slowly compared to the fermionic modes.
In this formulation, the functional space of gauge connections replaces spacetime
as the natural arena for the theory, and the long-distance behavior is encoded in
quantized Berry phases associated with the infrared clouds. Our results suggest
that the infrared sector of QCD exhibits features reminiscent of a
\emph{topological phase}, similar to those encountered in condensed-matter
systems, where topological protection replaces dynamical confinement at low
energies. In this geometric framework, color-neutral composites such as
quark--gluon and gluon--gluon clouds arise as topological bound states described
by functional holonomies. Illustrative applications to hadronic excitations are
discussed within this approach, including mesonic and baryonic examples.
This perspective provides a unified picture of infrared dressing and topological
quantization, establishing a natural bridge between non-Abelian gauge theory,
adiabatic Berry phases, and the topology of the space of gauge configurations.
}
\begin{document} 
\maketitle
\flushbottom

\section{Introduction}

The study of QCD in the low-energy regime constitutes one of the most challenging problems in modern physics. Among its many difficulties, its nonperturbative character is perhaps the most distinctive, since --in contrast with quantum electrodynamics (QED) in the infrared—it highlights the nonlinear nature of the theory and leads to additional subtleties that make its understanding even more demanding.

However, proposing a consistent approach to this problem is not straightforward unless one follows certain analogies inspired by QED. In fact, the infrared problem in QED was first addressed by Bloch and Nordsieck \cite{BN}, who proposed that the correct prescription for handling infrared divergences is to sum over all possible final coherent states of soft photons, assuming that the finite sensitivity of any detector provides a natural infrared cutoff. This physical idea was later generalized by Kinoshita \cite{KI}, Lee, and Nauenberg \cite{LN}, and systematically reformulated by Yennie, Frautschi, and Suura \cite{YFS,weinberg}.

Although Bloch and Nordsieck did not regard their construction as a regularization in the modern sense, their reasoning effectively anticipated that interpretation: the detector resolution acts as a physical infrared regulator, ensuring finite transition probabilities.

In the 1960s and 1970s --possibly inspired by the emerging rigorous approaches to quantum field theory-- the treatment of infrared divergences began to be addressed in a genuinely nonperturbative context~\cite{summers}. Within this framework, Chung~\cite{chung}, Kibble~\cite{kibble1,kibble2,kibble3,kibble4}, Kulish, and Faddeev~\cite{KF} (CKKF) proposed the interpretation of electron–photon clouds as the elementary physical entities of QED. Since then, several rigorous studies of the nonperturbative QED vacuum have been carried out \cite{FS,MorchioStrocchi,StrocchiBook,MSTR,ref1,Buchholz1,Buchholz2,Herdegen}.

The CKKF dressing is a mathematically suggestive object because it resembles a Wilson loop, and under very general conditions the dressed state possesses a well-defined quantum statistics. Therefore, if the physical problem is addressed appropriately, this framework may offer new interpretative insights.

In the infrared treatment of QED within the adiabatic approximation \cite{JG000,JG01}, it is remarkable that the CKKF dressed states can not only be constructed explicitly, but also that the entire set of IR–QED states is \emph{discrete} \cite{GM}. This discreteness arises because the functional Berry phase
\[
\Phi[C]=\oint_C \mathcal{A}
\]
which dresses the \emph{in} and \emph{out} spin-$\tfrac{1}{2}$ states, is quantized.

This implies the important physical fact that the electron-photon cloud is a coherent composite state, namely, a global configuration that allows one to interpret the infrared sector of QED as a \emph{topological phase}, closely analogous to the topological phases familiar in modern condensed--matter physics \cite{Hasan,Qi,Moore,Kane,Berne}.

In IR-QCD, however, the construction is more intricate owing to the nonlinear nature of gluons: one must separate soft from hard gluons and account for the fact that quarks are surrounded by a cloud of soft gluons. In this description,
it is reasonable to assume that hard gluons are absorbed into the dressed states --since they cannot remain free due to color confinement-- while the existence of two distinct Berry connections \cite{Berry} suggests that the corresponding clouds commute with each other.

In this paper, we extend the ideas previously developed for IR--QED to the
non-Abelian case of QCD. We begin by analyzing how quark--gluon and
gluon--gluon clouds emerge simultaneously in the dressed \emph{in} and
\emph{out} states. This simultaneous dressing breaks the standard Fock-space
structure of asymptotic states, leading instead to a decomposition of the
Hilbert space into topological or superselection sectors characterized by
quantized infrared fluxes.

From this viewpoint, the infrared phase of QCD can be regarded as a collective
topological regime in which color and gauge degrees of freedom become
inseparable. The corresponding dressed states carry nontrivial Berry
holonomies associated with both the fermionic and gluonic connections,
providing a natural generalization of the CKKF construction to the
non-Abelian case.

Beyond its formal consistency, this approach offers additional conceptual
advantages. It provides a natural extension of the notion of charged states
introduced by Morchio and Strocchi \cite{MSTR}, as well as of the algebraic framework
developed by Buchholz \cite{Buchholz1,Buchholz2}, both of which, although originally conceived for
nonperturbative QED, are expected to play a crucial role in understanding
color confinement and the possible emergence of a mass gap.

The paper is organized as follows. In Sec.~II, we discuss QCD in the infrared regime and show how this regime can be identified as a new phase of QCD, analogous to the topological phases known in condensed matter physics. In Sec.~III, we explain the results obtained using physical arguments and show how the quantization of the flux emerges as a consequence of the statistics of the infrared clouds. In Sec.~IV, we present four applications of the approach --to the neutral pion, the proton, the tetraquark and the pentaquark -- which, although they are bound states, can be consistently studied from the functional point of view. Section~V summarizes the main conclusions of the approach.  Finally, the appendix presents a didactic example that illustrates the central idea of the approach.

\section{The Infrared Regime of QCD as a Topological Phase}

In this section, we explore the infrared regime of QED and QCD as a
\emph{topological phase}, formally reminiscent of the topological phases
encountered in condensed-matter systems \cite{Hasan,Qi,Moore,Kane,Berne,witten01}.   
The analogy, however, is only structural: in gauge theories, the physical
origin of topological phases lies in the identification of CKKF dressed states and in the quantization
of \emph{functional} fluxes, which determine the statistics and topology of
the dressed \emph{in} and \emph{out} states.

To make these ideas more concrete, we consider QCD with \emph{massless} fermions. 
This assumption is not essential but convenient, as it removes the explicit dynamical mass scale 
and allows us to isolate purely geometric (topological) contributions.
In the infrared regime—where the adiabatic approximation holds—the dynamical
phases associated with massive excitations become irrelevant, while the
geometric phases encoded in the fermionic determinant remain as the essential
nonperturbative information.

Under these premises, our starting point is the generating functional
\begin{align}
Z &= \int [{\cal D}A]\,{\cal D}\bar\psi\,{\cal D}\psi~
\exp\!\left[i\int d^4x\left(
-\frac{1}{4}\,\mathrm{tr}\left(F_{\mu\nu}F^{\mu\nu}\right)
+\bar\psi\,i\slashed{D}[A]\,\psi\right)\right] \nonumber\\[4pt]
&= \int [{\cal D}A]~
e^{i\int d^4x\,-\tfrac{1}{4}\,\mathrm{tr}(F_{\mu\nu}F^{\mu\nu})}\,
\det\!\big(i\slashed{D}[A]\big),
\label{FG}
\end{align}
where $[{\cal D}A]$ denotes the Faddeev--Popov measure.

To evaluate the fermionic determinant in the adiabatic approximation, we assume that the background gauge
field varies slowly with the Minkowski time variable $t$.
For each fixed $t$ we solve the instantaneous Dirac eigenvalue problem
\begin{equation}
H_D(t)\,\varphi_m({\bf x};t) = E_m(t)\,\varphi_m({\bf x};t),
\end{equation}
where $H_D(t)= -i\gamma^0 \boldsymbol{\gamma}
\cdot (\nabla - i\mathbf{A})$ and
$\{\varphi_m\}$ form an orthonormal basis.  
For massless fermions, the spectrum is symmetric under
$E_m \leftrightarrow -E_m$, so the net dynamical phase cancels between
conjugate pairs of modes.

The fermionic fields are expanded as
\begin{align}
\psi(t,{\bf x}) &= \sum_m a_m(t)\,\varphi_m({\bf x};t),\\
\bar{\psi}(t,{\bf x}) &= \sum_m \bar{a}_m(t)\,\varphi_m^\dagger({\bf x};t).
\end{align}
Substituting into the action gives \footnote{This procedure was developed in Ref.~\cite{CGL}.}
\begin{equation}
S_F = \int dt\,\sum_{m,n}
\bar{a}_m(t)\Big[
  i\delta_{mn}\partial_t
  - i\mathcal{A}_{mn}(t)
  - E_m(t)\delta_{mn}
\Big]a_n(t),
\end{equation}
with the (anti-Hermitian) Berry connection
\begin{equation}
\mathcal{A}_{mn}(t)
   = i\,\langle \varphi_m | \partial_t | \varphi_n \rangle.
\end{equation}
In the chiral (massless) limit, the Berry connection acts nontrivially within
the degenerate subspaces of positive and negative energy modes, and thus
encodes the purely geometric evolution of the fermionic vacuum .

Performing the Grassmann path integral over the coefficients $a_m$ yields
\begin{equation}
\det(i\partial_t - E - i\mathcal{A})
   = \exp\!\Big[i\,\mathrm{Tr}\,
   \ln(i\partial_t - E - i\mathcal{A})\Big].
\end{equation}
In the adiabatic limit, where $\mathcal{A}$ varies slowly compared to
the instantaneous spectrum $E_m(t)$, the determinant reduces to
\begin{equation}
\det(i\slashed{D})
   \;\approx\;
   \exp\!\left[-i\int dt\, \sum_m E_m(t)\right]\,
   \mathrm{Tr}\,\mathcal{P}
   \exp\!\left(i\oint_C \mathcal{A}_F\right).
\end{equation}
For massless fermions, the first (dynamical) exponential cancels between
$\pm E_m$ pairs, while the second term—representing the holonomy of the
Berry connection—survives as the \emph{topological} contribution%
\footnote{Henceforth we shall set the dynamical phase $E_n$ to zero.}.
In other words,
\begin{equation}
\det(i\slashed{D})
   \;\approx\;
   \mathrm{Tr}\,\mathcal{P}
   \exp\!\left(i\oint_C \mathcal{A}\right).
\end{equation}

This geometric phase reflects the parallel transport of the degenerate
eigenspaces and provides the nontrivial infrared (IR) structure of the theory.

It is instructive to compare this result with that discussed in~\cite{JG000,GM}, 
where the chiral anomaly and Fujikawa’s method play a central role~\cite{fujikawa1,fujikawa222}. 
First, note that Eq.~(9) is fully valid in the deep infrared regime, 
i.e., when the kinetic energy of the gauge bosons is essentially negligible,%
\footnote{This is analogous to the quantum-mechanical propagator 
$G(x,x';\Delta t)=\langle x|\,e^{-i\left(\frac{\hat{p}^2}{2m}+V(\hat{x})\right)\Delta t}\,|x'\rangle
\;\to\; e^{-i V(x)\Delta t}\,\delta(x-x')$ in the $m\!\to\!\infty$ limit. 
An explicit example is discussed in Appendix~A.}

Second, the apparent absence of a chiral anomaly is only illusory: since 
$\{\gamma_5,\slashed{D}\}=0$, the nonzero eigenvalues of $\slashed{D}$ occur in $\pm\lambda$ pairs and cancel in $\mathrm{Tr}\,\gamma_5$, whereas unpaired zero modes remain. Their chiral asymmetry defines the index \cite{atiyah1,atiyah2}
\[
\mathrm{index}(\slashed{D}) \equiv n_{+}(0)-n_{-}(0)
=\frac{e^2}{16\pi^2}\int d^4x\,F_{\mu\nu}\tilde F^{\mu\nu},
\]
so the anomaly is indeed present here as well.

The generating functional with the fermionic degrees of freedom integrated out becomes
\begin{equation}
Z = \langle {\rm out} | {\rm in} \rangle
   = \int [{\cal D}A]~\Delta_{\rm FP}[A]~
   \exp\!\left[i\!\int d^4x\,\frac{1}{4}\,
   \mathrm{tr}\!\left(F_{\mu\nu}F^{\mu\nu}\right)\right]
   \times \mathcal{P}\exp\!\left(i\!\oint_C {\cal A}_F\right),
\end{equation}
where the gauge has been fixed and the Faddeev--Popov determinant
$\Delta_{\rm FP}[A]$ is included.

The evaluation of the functional integral becomes considerably simpler in the
adiabatic approximation \footnote{In the present adiabatic treatment, the Faddeev--Popov determinant does not modify the effective action: in the Abelian case it is field--independent and can be absorbed in the overall normalization, while in the non--Abelian case it only contributes at the Gaussian level without altering the geometric (Berry--phase) structure. Moreover, since the background configurations are smooth and slowly varying, no Gribov ambiguities arise within this approximation.}, as only the quadratic contributions (from fluctuations,
gauge fixing, and ghosts) need to be retained.

Gathering all the contributions, the partition function can be written as
\begin{equation}
Z = \langle {\rm out} ~|~ {\rm in} \rangle
   = \mathcal{P}\exp\!\left\{\,i\!\oint_C
   \big(\mathcal{A}_G + \mathcal{A}_F\big)\right\}.
\end{equation}

We now define the corresponding operator as
\begin{equation}
    {\cal U}_C = \mathcal{P}\exp\!\left\{\,i\!\oint_C
    \big(\mathcal{A}_G + \mathcal{A}_F\big)\right\},
\end{equation}
so that
\begin{equation}
|{\rm out}\rangle = \mathcal{U}_C\,|{\rm in}\rangle,
\qquad\text{and hence}\qquad
Z = \langle {\rm out}|{\rm in}\rangle
  = \langle {\rm in}|\,\mathcal{U}_C\,|{\rm in}\rangle.
\end{equation}

When a gauge--invariant observable associated with the closed contour $C$ is
required (the holonomy around a closed path in configuration space), one takes
the trace over the internal fiber:
\begin{equation}
\mathcal{W}_C =
\mathrm{Tr}\,\mathcal{U}_C
=
\mathrm{Tr}\,\mathcal{P}\exp\!\left\{\,i\!\oint_C
\big(\mathcal{A}_G + \mathcal{A}_F\big)\right\},
\end{equation}
which represents the functional analogue of a Wilson loop, now containing both
the gluonic and fermionic Berry phases.
Note that $\mathrm{Tr}(\cdot)$ does not act on the external states
$|{\rm in/out}\rangle$, but rather converts the holonomy into a gauge--invariant
scalar observable.

\section{Interpretation Issues}

In this section, we examine in more detail a number of interpretational aspects that are crucial for the physical understanding of the present framework and that demand a careful level of theoretical precision.

\subsection{Quantization of Functional Flux and the Statistical Nature of Dressed Clouds}

The product of holonomies, 
\(
\mathcal{U}_C = U_G(C)\,U_F(C)
= \mathcal{P}\exp\!\left[i\!\oint_C
(\mathcal{A}_G + \mathcal{A}_F)\right],
\)
represents the fermionic dressed state of the theory --the coherent quark state surrounded by its soft gluonic cloud. Although the dressing contains bosonic degrees of freedom, the combined holonomy remains fermionic because the quantized fermionic flux determines the overall parity of the total phase. In this sense, the product of gluonic and fermionic
holonomies encodes the geometric entanglement between matter and gauge fields, generalizing the CKKF dressing to the non-Abelian infrared regime.

This fact implies that the phase is quantized according to the following rule:
\begin{equation}
e^{\,i\!\oint_C (\mathcal{A}_G + \mathcal{A}_F)} = (-1)^{n_F}(+1)^{n_G}
           = (-1)^{n_F + 2n_G},
\end{equation}
so that
\begin{equation}
\oint_C (\mathcal{A}_G + \mathcal{A}_F)
          = \pi n_F + 2\pi n_G
          \;\equiv\; \pi n_F
          \quad (\mathrm{mod}\;2\pi).
\end{equation}
The accumulated phase under adiabatic functional transport is therefore
quantized in multiples of~$\pi$: even values of $n_F$ correspond to bosonic
sectors ($+1$), while odd values correspond to fermionic sectors ($-1$).
This quantization condition determines the \emph{statistics} of the dressed
cloud rather than restricting the possible spin content of the physical
states. In other words, the discrete Berry phases classify the topological
sectors of the infrared vacuum (fermionic or bosonic), but they do not prevent
the construction of states with different total spin or angular momentum.
\subsection{Infrared Clouds and Topological Structure}

In IR--QED the nature of the dressed cloud already lies outside the standard
relativistic interpretation of electrodynamics. Physical asymptotic states are
not bare electrons plus radiation, but electron states coherently dressed by a
soft photon cloud. In the adiabatic infrared regime, the electron and its cloud
behave as a single composite object, effectively labelled by a given functional
(topological) sector. The soft electromagnetic radiation is not absent, but its
infrared component is absorbed into the very definition of the asymptotic
states.

In IR--QCD the situation is qualitatively different. Because QCD is a
nonlinear, non-Abelian gauge theory, two types of infrared clouds coexist:
quark--gluon and gluon--gluon. In the adiabatic regime, these are naturally
described by two distinct Berry connections, $\mathcal{A}_F$ and
$\mathcal{A}_G$, acting on different fibers of the state space. Their holonomies define two simultaneously quantized functional fluxes which, while subject to global constraints such as Gauss's law and center neutrality, remain topologically distinguishable. The product of these holonomies encodes the fully dressed fermionic state in QCD, and leads to a  pattern of infrared superselection sectors than in the Abelian case.
\subsection{Composition of the Infrared States}
Given these two features --namely, that the infrared clouds in QCD are
topologically independent and that the quark-- gluon clouds carry a fermionic
(half-integer) statistical phase --it appears natural to expect the formation
of bound configurations. Such composite states may exhibit either
half-integer or integer spin, depending on how the fermionic and gluonic
clouds combine. In particular, fermionic clouds can merge to produce
bosonic configurations (as in meson-like structures), while the attachment of
gluonic clouds to fermionic ones preserves a half-integer character, resembling baryonic or hybrid excitations. 

In this way, the quantization of
functional flux and the statistical structure of the infrared clouds
provide a natural kinematical origin for the emergence of bound states in
QCD.

\section{Applications}

As an application of these ideas, we now consider the superposition of
quantized functional fluxes as the building blocks of physically realized
states. In other words, since we are assuming that IR--QCD can be viewed as a
topological phase, it is reasonable to admit that, in the infrared, the role
of spacetime is effectively replaced by the space of functionals
$\mathcal{U}_{F,G}$---the space of fermionic and gluonic holonomies. Within this framework, the fundamental degrees of freedom are not local field  excitations but global, quantized configurations of functional flux—essentially, a texture. 
By superposing these configurations, one can construct infrared hadronic states whose properties—spin, statistics, and color neutrality—emerge from the  topological structure of the underlying functional space.

\subsection{The Neutral Pion as a Topological Composite State}

Let's consider the neutral pion  
in the functional framework developed above.  The neutral pion can be described
as a composite configuration of two conjugate functional fluxes: one
fermionic (quark--gluon) and one antifermionic (antiquark--gluon). Each
constituent is represented by a holonomy
$U_C(\pm\tfrac{1}{2},1)$, where the sign of the fermionic flux corresponds to
the quark or antiquark sector, while the gluonic flux $n_G=1$ provides a
common topological background:
\[
|\pi^0\rangle \;\sim\;
\Big(
U_C(+\tfrac{1}{2},1)
\;\oplus\;
U_C(-\tfrac{1}{2},1)
\Big)\,|0\rangle.
\]
The conjugate fermionic phases, $+\pi$ and $-\pi$, cancel in the total
holonomy, producing a bosonic and topologically neutral configuration
stabilized by the gluonic background.

Group-theoretically, this construction corresponds to the tensor product
\[
\mathbf{3} \otimes \bar{\mathbf{3}}
\;=\;
\mathbf{8} \;\oplus\; \mathbf{1},
\]
where the singlet $\mathbf{1}$ represents the colorless, gauge-invariant
component of the pion. The adjoint $\mathbf{8}$ corresponds to the non-singlet
color excitations, which are suppressed in the infrared due to topological
neutrality.

In the functional representation, the pion wave functional can thus be viewed
as a symmetric combination of two conjugate holonomies,
\[
\Psi_{\pi^0}[\mathcal{U}_1,\mathcal{U}_2]
\;=\;
\frac{1}{\sqrt{2}}\,
\Big(
f(\mathcal{U}_1,\mathcal{U}_2)
\;+\;
f(\mathcal{U}_2,\mathcal{U}_1)
\Big),
\]
where $f$ is a functional of the individual holonomies in the space
$\mathcal{U}_{F,G}$. This symmetric structure enforces both bosonic statistics
and color neutrality, analogously to the singlet projection in
$\mathbf{3}\otimes\bar{\mathbf{3}}$.

Therefore, the neutral pion emerges as a \emph{topological composite state} in
the functional space $\mathcal{U}_{F,G}$, built from two conjugate quantized
infrared holonomies. Its color neutrality and bosonic nature are direct
consequences of the functional flux quantization, rather than of any explicit
dynamical potential or confinement mechanism.
In a sense, this construction provides a geometric ``shortcut'' to the infrared spectrum: it bypasses the dynamical problem while still capturing its physical consequences. The existence of bound, colorless states thus 
appears as a topological necessity rather than as the outcome of a specific potential or interaction.
\subsection{The Proton as a Topological Composite}

From the functional point of view, the proton can be regarded as a composite
configuration built from three fermionic functional fluxes, each corresponding
to a quark--gluon cloud, coupled to a common gluonic background:
\[
U_p(C)
\;\sim\;
U_C\!\left(\tfrac{1}{2},1\right)
\otimes
U_C\!\left(\tfrac{1}{2},1\right)
\otimes
U_C\!\left(\tfrac{1}{2},1\right).
\]
In the group-theoretic language, each $U_C(\tfrac{1}{2},1)$ transforms in the
fundamental representation $\mathbf{3}$ of $SU(3)$, while the gluonic component
belongs to the adjoint representation $\mathbf{8}$ or its topological
projection in the infrared regime. Consequently, the proton corresponds to the
triple tensor product of three fundamental representations,
\[
\mathbf{3} \otimes \mathbf{3} \otimes \mathbf{3}
\;=\;
\mathbf{10} \;\oplus\; \mathbf{8} \;\oplus\; \mathbf{8} \;\oplus\; \mathbf{1},
\]
whose color-singlet component $\mathbf{1}$ represents the physically
observable, gauge-invariant proton state. This singlet corresponds to the
completely antisymmetric combination of three color indices, ensuring total
color neutrality.

In the functional representation, the proton wave functional can be written as
an antisymmetric combination of three holonomies,
\[
\Psi_p[\mathcal{U}_1,\mathcal{U}_2,\mathcal{U}_3]
\;=\;
\frac{1}{\sqrt{6}}
\sum_{\sigma\in S_3}
\mathrm{sign}(\sigma)\,
f(\mathcal{U}_{\sigma(1)},\mathcal{U}_{\sigma(2)},\mathcal{U}_{\sigma(3)}),
\]
where $f$ is a symmetric functional of the individual holonomies
$\mathcal{U}_i \in \mathcal{U}_{F,G}$ and $S_3$ denotes the permutation group
acting on the three fermionic fluxes. The antisymmetrization over $S_3$ plays
the same role as the Levi-Civita tensor $\epsilon_{abc}$ in conventional color
space: it projects the functional configuration onto the singlet sector of
$SU(3)$.

This representation-theoretic structure shows that the proton state in the
functional framework corresponds to the colorless, totally antisymmetric
combination of three fermionic holonomies. The geometric content of this
construction is that the proton arises as a topologically stable,
gauge-invariant configuration in the functional space $\mathcal{U}_{F,G}$,
where the antisymmetry condition enforces both color neutrality and fermionic
statistics. Thus, the familiar algebraic structure of the baryon octet and
decuplet finds a natural interpretation as the decomposition of tensor products
of quantized infrared holonomies.

\subsection{The Tetraquark as Topological Composite}

In the present framework, hadronic bound states are regarded as \emph{topological composites} rather than as Fock superpositions of free quarks and gluons. Each quark–gluon cloud is represented by a Wilson-type holonomy in the functional configuration space $\mathcal{A}/\mathcal{G}$,
\begin{equation}
U_C(R) \;=\; P\,\exp\!\left(i\oint_C \mathcal{A}[A]\right),
\label{eq:holonomy_single_tetra}
\end{equation}
where $\mathcal{A}[A]$ denotes the Berry functional connection associated with the adiabatic evolution of the gauge field. Physical states correspond to closed holonomies, whose nontrivial topology in $\mathcal{A}/\mathcal{G}$ encodes the infrared dressing of the theory.

A tetraquark, composed of two quarks and two antiquarks, can thus be expressed as a functional composite of four intertwined holonomies,
\begin{equation}
U_T \;\sim\;
U_{C_1}\!\left(\tfrac{1}{2}\right)
\otimes U_{C_2}\!\left(\tfrac{1}{2}\right)
\otimes U_{\bar{C}_1}\!\left(\tfrac{1}{2}\right)
\otimes U_{\bar{C}_2}\!\left(\tfrac{1}{2}\right),
\label{eq:tetra_tensor_simple}
\end{equation}
where each factor corresponds to a quark or antiquark cloud evolving adiabatically along a path $C_i$ in the functional space.  
The combined trajectory $\Gamma_T$ defines a \emph{non-factorizable holonomy},
\begin{equation}
U_T \;=\; {\rm Tr}\!\left[
P\,\exp\!\left(i \oint_{\Gamma_T} \mathcal{A}\right)
\right],
\label{eq:tetra_path_simple}
\end{equation}
representing a closed network of curves in $\mathcal{A}/\mathcal{G}$ linked through their common gauge connection. 
This composite holonomy encodes the topological correlations between the constituent clouds, providing a geometric realization of confinement as a condition of topological closure rather than as a potential interaction.

From the representation-theoretic point of view, the tensor product of two quarks and two antiquarks can be decomposed as
\begin{equation}
(3\!\otimes\!3)\!\otimes\!(\bar{3}\!\otimes\!\bar{3})
\;=\;
(6\oplus\bar{3}) \!\otimes\! (\bar{6}\oplus 3)
\;=\;
1 \oplus 8 \oplus 8 \oplus 10 \oplus \overline{10} \oplus 27,
\label{eq:rep_sum_tetra}
\end{equation}
where the color singlet corresponds to the physically confined configuration.  
In our functional formulation, each direct–sum component in~\eqref{eq:rep_sum_tetra} represents a distinct \emph{topological sector} of the composite holonomy, characterized by the pattern of linking between its constituent loops.

Geometrically, the tetraquark can thus be visualized as a closed network of four functional paths $\{C_i\}$ whose linking structure defines a nontrivial topology in $\mathcal{A}/\mathcal{G}$.  
The corresponding holonomy~\eqref{eq:tetra_path_simple} acts as a topological order parameter distinguishing different composite configurations, in close analogy with non-Abelian anyonic systems in condensed matter.  
In this picture, the tetraquark emerges as a \emph{topological molecule} in functional space, where the binding arises from the entanglement of holonomies rather than from local gauge potentials.

\subsection{The Pentaquark as Topological Composite}

From the functional point of view, a pentaquark can be regarded as a composite
configuration built from four fermionic fluxes and one antifermionic flux,
each represented by a functional holonomy coupled to a common gluonic background:
\[
U_P(C)
\;\sim\;
U_C\!\left(\tfrac{1}{2},1\right)
\otimes
U_C\!\left(\tfrac{1}{2},1\right)
\otimes
U_C\!\left(\tfrac{1}{2},1\right)
\otimes
U_C\!\left(\tfrac{1}{2},1\right)
\otimes
U_C\!\left(\tfrac{1}{2},\bar{1}\right),
\]
where each $U_C(\tfrac{1}{2},1)$ corresponds to a quark--gluon cloud in the
fundamental representation $\mathbf{3}$ of $SU(3)$, and
$U_C(\tfrac{1}{2},\bar{1})$ represents the corresponding antiquark cloud in the
antifundamental representation $\bar{\mathbf{3}}$.
The overall color structure follows from the tensor product
\[
\mathbf{3} \otimes \mathbf{3} \otimes \mathbf{3} \otimes \mathbf{3} \otimes \bar{\mathbf{3}}
\;=\;
\mathbf{1} \;\oplus\; 4\,\mathbf{8} \;\oplus\; 3\,\mathbf{10} \;\oplus\; \cdots,
\]
which indeed contains a color-singlet component $\mathbf{1}$ corresponding to
the gauge-invariant, observable pentaquark configuration.

In the functional representation, the pentaquark wave functional takes the schematic form
\[
\Psi_P[\mathcal{U}_1,\ldots,\mathcal{U}_5]
\;=\;
\mathcal{A}\!\left[
f(\mathcal{U}_1,\mathcal{U}_2,\mathcal{U}_3,\mathcal{U}_4,\mathcal{U}_5)
\right],
\]
where $\mathcal{A}$ denotes the antisymmetrization operator acting over the
four fermionic holonomies in the fundamental representation and the single
antifermionic holonomy, ensuring total color neutrality.
As in the baryonic case, the antisymmetrization enforces the projection onto
the singlet sector of $SU(3)$, while the functional dependence of $f$
encodes the spatial and gauge correlations among the constituent clouds.

Physically, this construction represents a higher-order topological configuration in the functional space $\mathcal{U}_{F,G}$, where the four fermionic and one antifermionic fluxes are bound through a common infrared gluonic background. From the geometric point of view, the pentaquark appears as a multi-linked configuration of functional holonomies, whose topological stability follows from the quantization of the non-Abelian Berry flux.

In this sense, both baryons and exotic multiquark states such as tetraquarks and pentaquarks arise as distinct topological sectors of the same underlying functional geometry, differing not by their microscopic composition but by the linking structure of their quantized flux loops.

\section{Discussion of the Approach}

We have argued that the infrared sector of QCD can be understood, within the
adiabatic approximation, as a \emph{topological phase}—closely analogous to the
topological phases encountered in condensed-matter systems. In those systems,
topological protection is typically realized at the boundary of a bulk medium;
in our case, spacetime itself is effectively replaced by the functional space
of gauge configurations, and the role of the boundary is played by the
quantized functional Berry phases associated with the infrared clouds. In QCD
these clouds appear in two distinct forms—quark–gluon and gluon–gluon—marking a
fundamental departure from the Abelian case.

In Abelian gauge theories, where only a single type of infrared cloud exists
and the basic degrees of freedom are the electron and the photon, both the
theoretical structure and the experimental signatures are comparatively
restricted. Observable effects are expected only in the deep infrared, such as
minute deviations from the Rayleigh–Jeans law or anomalies in the low-frequency
tail of the CMB spectrum.

By contrast, the non-Abelian character of infrared QCD, with its inherent
nonlinearity and coexistence of multiple cloud species, opens an entirely new
set of conceptual and physical possibilities. Hadronic excitations can be
studied without invoking an explicit confinement potential, since topological
protection now operates in the \emph{space of fields} rather than in spacetime
itself. This geometrical perspective provides a natural setting in which
color-neutral bound configurations emerge as topologically stable composites,
thus reframing—rather than solving—the confinement problem within a functional,
topological context.

Finally, this construction connects naturally with the framework of asymptotic
symmetries~\cite{stro1,stro2,stro3}. In that setting, the celestial sphere
appears as the boundary $S^2$ at null infinity, while in our approach it
corresponds to a specific topological sector of the functional space, where
infrared configurations are classified by quantized holonomies (or functional
fluxes) rather than by local field excitations. This correspondence builds a
conceptual bridge between the infrared memory effects described in the
asymptotic formalism and the Berry-phase topology emerging in the adiabatic
regime.

\appendix
\section*{Appendix A: Low--Energy Limit of a Charged Particle  in a Dirac--Monopole Field}

To illustrate the origin of the topological dressing in a simple quantum--mechanical
setting, let us consider an electron moving in the background of a Dirac monopole
of magnetic charge $g$. The Hamiltonian is
\begin{equation}
H = \frac{1}{2m}(\mathbf{p} - e\,\mathbf{A}_M)^2,
\qquad
\nabla\times \mathbf{A}_M = g\,\frac{\mathbf{r}}{r^3},
\end{equation}
with the Dirac quantization condition $e g = 2\pi n$, $n\in\mathbb{Z}$.
The corresponding path integral for the quantum propagator reads
\begin{equation}
K(\mathbf{x}_f,\mathbf{x}_i;T)
= \int_{\mathbf{x}(0)=\mathbf{x}_i}^{\mathbf{x}(T)=\mathbf{x}_f}
\!\!\mathcal{D}\mathbf{x}(t)\;
\exp\!\left\{ i\!\int_0^T\!dt
\left[\frac{m}{2}\dot{\mathbf{x}}^2
+ e\,\dot{\mathbf{x}}\!\cdot\!\mathbf{A}_M(\mathbf{x})\right]\right\}.
\label{eq:Kmonopole}
\end{equation}

In the adiabatic (low--energy) regime, where the kinetic energy of the electron
is negligible compared with the magnetic background, the path integral localizes
at fixed spatial points. The propagator then reduces to
\begin{equation}
K(\mathbf{x}_f,\mathbf{x}_i;T)
\simeq
e^{\,i e\int_{\gamma_{\mathbf{x}_i,\mathbf{x}_f}}
\!\!\mathbf{A}_M\!\cdot d\mathbf{x}}\,
\delta^{(3)}(\mathbf{x}_f-\mathbf{x}_i),
\qquad
|\mathbf{p}|\to 0.
\label{eq:KlowE}
\end{equation}
The exponential factor is a purely geometric phase---the Wilson line of the
monopole potential---representing a \emph{topological dressing} of the electron state.

When the particle is transported around a closed loop $C$ on the sphere $S^2$,
\begin{equation}
\mathcal{U}[C]
= \exp\!\left(i e\oint_C\!\mathbf{A}_M\!\cdot d\mathbf{x}\right)
= \exp\!\left(i e g\,\Omega[C]\right),
\label{eq:HolonomyMonopole}
\end{equation}
where $\Omega[C]$ is the solid angle subtended by $C$.
This holonomy corresponds to the first Chern number of the $U(1)$ bundle and
provides the quantum--mechanical realization of the Berry phase for the monopole
configuration.

Hence, in the infrared limit, the physical electron state acquires a multiplicative
phase,
\begin{equation}
|\psi_{\rm phys}\rangle
= e^{\,i e\int \mathbf{A}_M\!\cdot d\mathbf{x}}\,
|\psi_{\rm bare}\rangle,
\end{equation}
which plays the same role as the Kulish--Faddeev soft--photon dressing in
infrared QED. In both cases, the adiabatic transport of slow degrees of freedom
produces a geometric phase that ensures gauge--invariant, infrared--finite states.

As an incidental remark, taking the trace of $K(\mathbf{x}_f,\mathbf{x}_i;T)$ and
evaluating it in Euclidean time leads to the partition function,
\begin{equation}
Z(T)
\;=\;
\mathrm{Tr}\,K_E(T)
\;=\;
\int d^3x\, K_E(\mathbf{x},\mathbf{x};T)
\simeq
\int d^3x\,
\exp\!\left[-\,e \oint_{\gamma_{\mathbf{x}}}
\mathbf{A}_E\!\cdot d\mathbf{x}\right],
\qquad
|\mathbf{p}|\to 0,
\label{eq:ZlowE}
\end{equation}
where $\gamma_{\mathbf{x}}$ denotes a closed Euclidean loop based at
$\mathbf{x}$. In this limit, the phase factor becomes a real exponential and
the trace of the kernel acquires the meaning of a partition function governed
by the Euclidean holonomy of the background gauge field. This provides a simple
example of a \emph{topological quantum mechanical system} in the sense discussed
in~\cite{dunne}, where the dynamics is fully encoded in a topological
phase and the energy spectrum is determined by the underlying holonomy class.

\acknowledgments
\noindent 
This research was supported by DICYT (USACH), grant number 042531GR\_REG.

\end{document}